# Nanoparticle mobility over a surface as a probe for weak transient disordered peptide-peptide interactions

*Indrani Chakraborty [†], Gil Rahamim [‡], Ram Avinery [‡], Yael Roichman\* [†,‡] and Roy Beck\* [‡]*

[†]School of Chemistry, Tel Aviv University, Tel Aviv 6997801, Israel

[‡]School of Physics and Astronomy, Tel Aviv University, Tel Aviv 6997801, Israel

ABSTRACT: Weak interactions form the core basis of a vast number of biological processes, in particular, those involving intrinsically disordered proteins. Here, we establish a new technique capable of probing these weak interactions between synthetic unfolded polypeptides using a convenient yet efficient, quantitative method based on single particle tracking of peptide-coated gold nanoparticles over peptide-coated surfaces. We demonstrate that our technique is sensitive enough to observe the influence of a single amino acid mutation on the transient peptide-peptide interactions. Furthermore, the effects of buffer salinity, expected to alter weak electrostatic interactions, are also readily detected and examined in detail. The method presented here has the potential to evaluate in a high throughput manner, weak interactions for a wide range of disordered proteins, polypeptides, and other biomolecules.

The vast majority of biological functions rely on reversible interactions between biomolecules. These complex interactions depend on non-trivial combinations of parameters, which may include the structure of the biomolecules, distribution of charged and hydrophobic/hydrophilic monomers, as well as the presence of moieties capable of forming hydrogen and disulfide bonds. Variation in the relative strength of these interactions results in a broad spectrum of interaction modes between biomolecules. For example, strong, long-lived, specific interactions such as antigen-antibody interaction and DNA hybridization involve mainly hydrogen and disulfide bonds. In contrast, much weaker, transient, non-specific interactions are based on ionic bridging and steric repulsion.[1–3] While specific interactions have been studied and characterized extensively within the context of cell biology, pharmacology, and molecular biology, the characterization of non-specific interactions remains challenging. The key challenges are the transient nature of these bonds and the fact that the order of magnitude of the interaction energy is close to the thermal noise.

Intrinsically disordered proteins (IDPs) are a class of biomolecules whose function relies on weak interactions whose exact mechanisms are still largely unknown. IDPs, while retaining biological functionality, violate the conventional sequence-structure-function paradigm, since they display amino acid sequences that do not lead to singular, stable 3D structures.[4–8] These proteins are involved in a range of cellular functions, including transcription, translation, signaling, and regulation of protein assembly.[9–11] One such example is the neurofilament-tail proteins in neurons. Abnormality in the weak interactions of these IDP proteins may lead to neurodegenerative diseases like Alzheimer's, Charcot-Marie-Tooth, Parkinson's disease, and amyotrophic lateral sclerosis.[12–15] Other IDPs have also been shown to play an important role in a very broad spectrum of human diseases ranging from cancer to neurodegenerative diseases, infectious diseases, cardiovascular diseases and diabetes.[8,16,17] In recent years IDPs have also been observed to be important participants in liquid-liquid phase separation.[18–22] Therefore, it is of paramount importance to find suitable methods for studying and understanding these weak interactions.

A wide range of techniques exist to identify and characterize protein-protein interactions (PPIs), as summarized in recent reviews.[23–25] When choosing a method for analyzing protein-protein interactions, one should consider different parameters such as cost, time constraints, sample size requirement, energy resolution of the interactions, the environment in which the interactions take place (*in-vivo, in-vitro*, etc.), and structural interference (such as added labeling). Many conventional methods probe PPIs within a rather narrow range, and in a binary fashion - either the proteins bind, or they do not. Improved PPI resolution can be achieved using bulk, low-throughput spectroscopic techniques. Other powerful techniques such as atomic force microscopy (AFM)[26–28] and "tweezer"-based methods[29] acquire single-molecule PPI information, but are, by their very nature, low-throughput, yielding limited statistics. Other commonly used, yet low-throughput techniques include dual polarization interferometry,[30] proximity ligation assay,[31] tandem affinity purification,[32,33] analytical ultracentrifugation,[34] and surface force balance.[35]

In this paper, we suggest a new technique for probing weak interactions by characterizing the diffusion of peptide-coated gold nanoparticles (GNPs) near a surface coated with the same or an opposing peptide. In several previous works, a similar approach was used to study strong DNA-DNA interactions,[36] delocalized long-range polymer-surface interactions[37] and bulk-mediated diffusion on supported lipid bilayers.[38] Our approach is to tune experimental conditions to the regime in between the two extremes of arrested diffusion (due to strong interaction forces between probe particle and surface) and free diffusion (as is the case for purely electrostatic repulsion). Here, weak interactions exist between functionalized GNPs and glass surfaces that govern and modify the diffusive properties of these particles. For simplicity, we study weak interactions in disordered polypeptides with specific amino acid sequences designed by us. We show that by carefully analyzing the diffusive nature of peptide-coated GNPs near peptide-coated flat glass surfaces, we can detect the effect of a single mutation in the polypeptide sequence. We also explore the different factors that govern and control the weak peptide-peptide interactions and show that our technique is also sensitive to small changes in environmental or structural factors that influence these interactions. Even though this work is concerned with weak interactions in disordered polypeptides, we anticipate our method for analysis of weak pair-wise interactions to be applicable to a larger class of biomolecules.

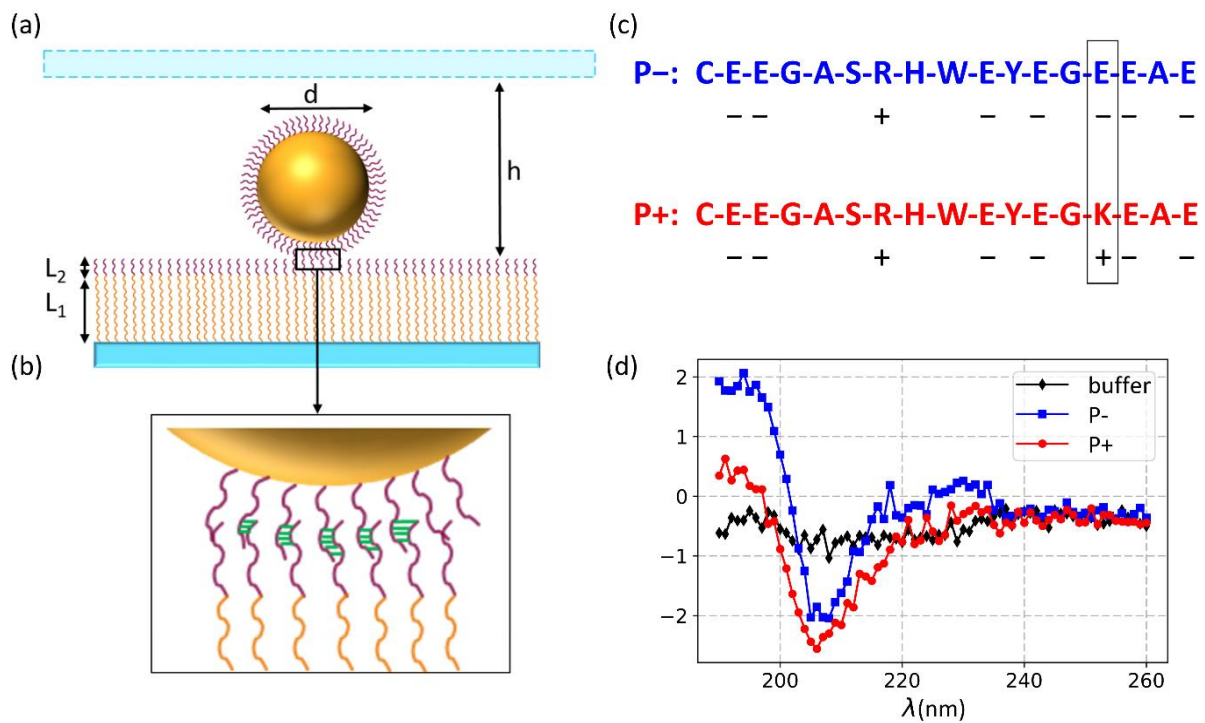

**Figure 1.** (a) Schematic diagram of the experimental setup, where (b) gives an illustration of the magnified view of the transient bonds (green lines) formed between the peptides on the GNPs and the peptides on the glass surface. The GNPs have a diameter of d = 40 nm, and the separation between the top and the bottom surfaces of the sample chamber (h) is about 21 µm. The peptides on the bottom surface are connected to the glass via a 10 kDa PEG linker molecule (shown in orange). In a fully extended state, the length of the PEG molecule is $L_1$ = 64 nm, and the length of the peptide is $L_2$ = 6 nm.[39] The two peptide sequences used in the experiment are given in (c). Note that the difference between the two sequences P− and P+ is a single mutation that is performed by replacing one glutamic acid (E) with lysine (K). The charge states of the constituent amino acids at pH = 7.5 are denoted below. (d) Circular dichroism measurements on the two peptides are typical of completely disordered peptides.[40]

In our experiments, we use a sample chamber with planar geometry in which a 21 µm thick liquid film containing GNPs (40 nm diameter) is sandwiched between two glass cover slips as shown in the schematic diagram of the experimental setup (Figure 1a). The peptides are covalently linked to the silane-PEG-maleimide coated bottom glass surface using thiol-maleimide bonds, whereas the top glass is passivated with PEG-silane. The presence of the PEG passivation layer prevents the GNPs from sticking to the glass due to non-specific Van der Waals forces. The formation of multiple weak bonds between the peptide-coated GNPs and the peptide-coated glass (Figure 1b) determines the nature of diffusion of the GNPs near the bottom glass wall. The diffusion of the GNPs is imaged with conventional dark-field microscopy (Olympus IX71), and the images are recorded at 50 frames per second (for further details see methods section). Several datasets of more than 10,000 trajectories each are recorded in each particular experiment. From these trajectories, only those longer than 60 frames are selected for calculating the mean square displacement (MSD) plot. Because of this large ensemble, we have sufficient statistics for calculating the different parameters of diffusion. The intrinsic geometry of the system, namely the small radius of the GNPs, ensures that only a few polypeptides on the GNPs come close to the peptides coated on the glass.

Having multiple bonds between the probe particle and the surface allows us to amplify the transient bond lifetime exponentially and to overcome the thermal noise level. Yet, the number of bonds is low enough to resolve the differences in binding affinity between polypeptides. A major benefit of our technique as compared to single particle techniques like optical-tweezers is the combination of acquiring information at the single particle level while also acquiring ensemble statistics.

We synthesized two peptide sequences, marked as P− and P+ by CEM Liberty Blue™ automated microwave peptide synthesizer and purified on a Waters AutoPurification system. Each of the sequences was composed of 17 amino acids. The two sequences differ only by a single amino acid (Figure 1c). When fully stretched, the polypeptide length is about 6 nm, which is 15% of the diameter of the GNPs.[39] We note that E14K mutation from P− to P+ changes the overall charge from −6.2e to −4.2e at pH =7.5.

Both P− and P+ are disordered peptides without any internal structure as indicated by circular dichroism measurements (Figure 1d).[40] The cysteine-containing peptides were coated onto the GNPs via the formation of gold-thiol bonds. The peptide coverage density on the GNPs was measured using quantitative fluorescence measurements (see methods and figure S1). The grafting density was observed to increase with added peptide concentration and saturate near a concentration of 100 μM. To avoid grafting ambiguity between GNP preparations, we chose conditions such that grafting densities were in the saturation region of Figure S1. The corresponding values of grafting density for P− and P+ were 0.48 ± 0.01 $nm^{−2}$ and 0.68 ± 0.01 $nm^{−2}$, respectively. AFM phase and topography images of the peptide-coated glass show homogenous surfaces without any indication for phase separation (see Figure S2). As another method to verify homogenous grafting, we coated cover-slips with a range of initial peptide concentrations within 0-100 μM and allowed the peptide-coated GNPs to diffuse on them and quantified their mobility. At low peptide coverage, we observed a higher number of slowly diffusing GNPs. As peptide coverage increased, the population of barely mobile GNPs decreased, indicating better passivation of the surface (Figure S3a-b).

We begin by observing the diffusion of GNPs coated with a given peptide (either P+ or P−) near a cover slip surface coated with the same peptide. We refer to this "self-interaction" mode as P+ on P+ and P− on P−, for simplicity. Three major classes of trajectories were observed in a series of different experiments: a) trajectories where the motion of the particles is confined to a very small area, that is the particles are 'immobile' b) trajectories where the particles perform an unrestricted random walk, and c) trajectories where the particles alternate between 'sticking' and 'hopping' (see Figure 2a). In all our subsequent analysis (except Figure S4), we exclude the 'immobile' particles, which are stuck from the beginning of the experiment. Namely, those particles which do not move more than a total distance of 1.2 µm in either the x or y directions in the first 400 ms of the recorded image sequences. The remaining mobile particles are included in further analysis. To analyze the intermittent sticking and hopping in the trajectories, we define a quantity called the sticking time $t_{st}$ which gives us the average time for each particle in which it moves less than an arbitrary cut-off of 0.6 µm in the x or y directions. From the probability distributions $P(t_{st})$ of sticking times (Figure 2b) we see that for P− on P−, $P(t_{st})$ is peaked about zero with a sharp fall in a time-span of 2 s. For P+ on P+, $P(t_{st})$ has a long tail extending up to ~ 25 s. This suggests that the interaction between P+ and P+ is much stronger than the interaction between P− and P−. With the addition of salt (160 mM NaCl), the interactions become stronger for both P− on P− and P+ on P+ leading to higher fractions of particles with larger $t_{st}$ (Figure 2c).

Since the diffusion of the individual probe particles can be anomalous, we define the mean square displacement (MSD) as $\langle \Delta x^2(\tau) \rangle = 4\widetilde{D}\tau^n$. Here, $\widetilde{D}$ is the transport coefficient, $n$ is the diffusion exponent, and $\tau$ is the lag time. Other than the sticking time distributions, a measurement of the fractions of mobile and immobile particles, as well as fractions of 'slow' ($\widetilde{D} < 1 \ \mu m^2/s^n$) and 'fast' ($\widetilde{D} > 1 \ \mu m^2/s^n$) trajectories for the two systems indicate the relatively stronger interactions for P+ on P+ (Figure S4). To examine this in detail, we analyzed a set of four different quantities to explore the essential differences in the diffusive behavior of P− on P− and P+ on P+, which are: a) the time ensemble averaged MSDs, b) the probability distribution of the diffusion exponents $P(n)$, c) the probability distribution of the probe particle displacements $G(\Delta x)$, and d) the probability distribution of the transport coefficients $P(\widetilde{D})$.

In practice, the diffusion exponent $n$ signifies the degree of anomaly of the diffusion of the probe particles. $P(n)$ is the probability distribution obtained by making linear fits to the individual MSDs of each particle in a log-log plot and calculating the slope $n$ of each fit. A pure diffusion would be indicated by $n$ values being closer to 1.0 while $n < 1.0$ would indicate subdiffusion such as those in 'stick and hop' trajectories. The probability distribution of displacements $G(\Delta x)$ is directly measured from the particle trajectories at a lag time of 0.06 s and it captures the Gaussian or non-Gaussian nature of the diffusion. Lastly, we analyze the probability distribution of the transport coefficients $P(\widetilde{D})$ instead of the diffusion constants of the individual particles since a fraction of the particles undergo anomalous diffusion. The transport coefficients are the exponentials of the intercepts of the individual MSDs in the log-log plot divided by a factor 4 for 2D diffusion. The four quantities calculated for P− on P− and P+ on P+ are shown in Figure 3.

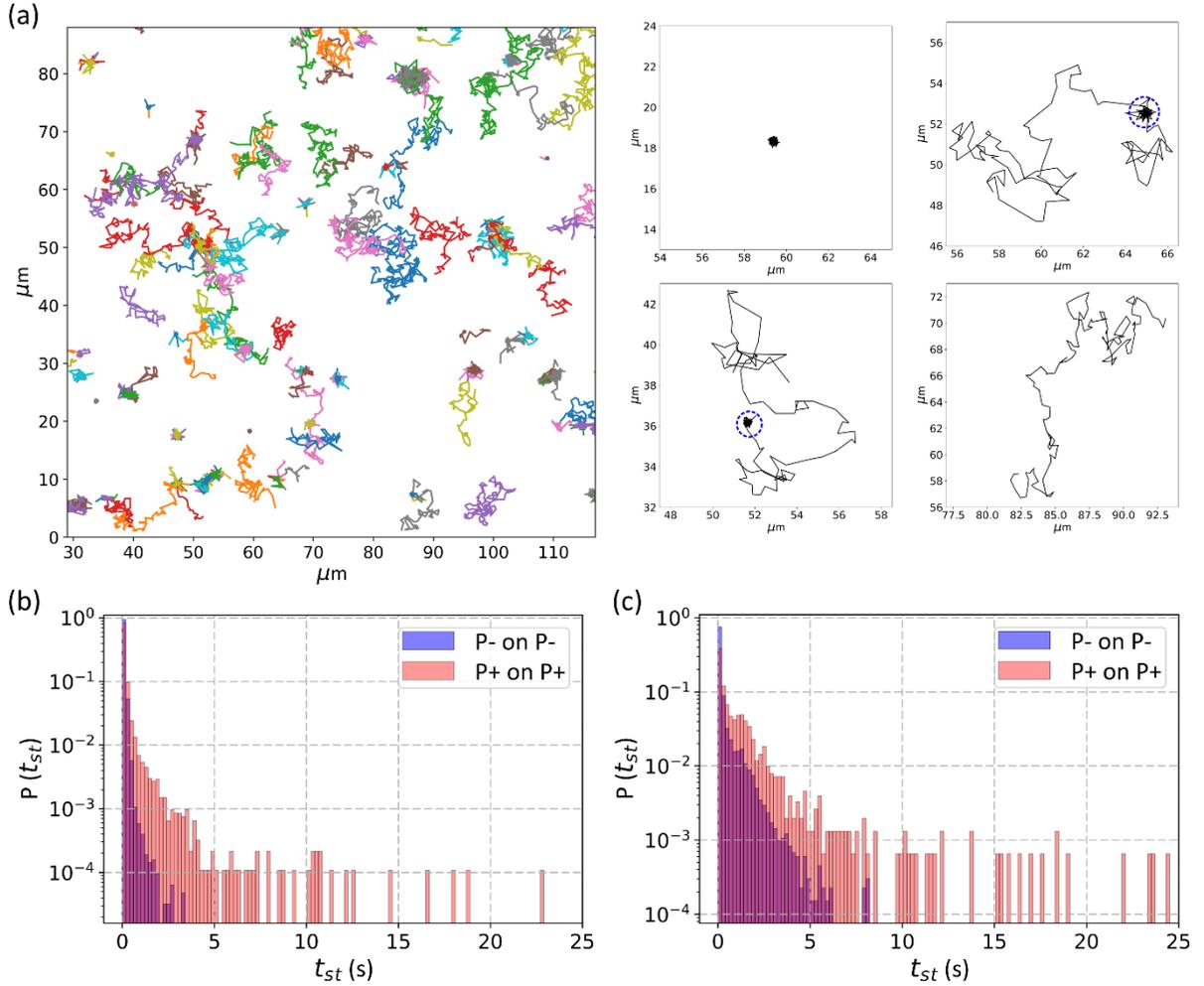

**Figure 2.** (a) Representative trajectories of peptide-coated GNPs on a peptide coated glass surface. A few example trajectories are shown on the right, including completely immobile, completely mobile, and particles which undergo sticking and hopping. For hopping particles, the sticking regions are indicated by blue dashed circles. (b) The probability distributions of sticking time of the GNPs show that the P+ to P+ interaction is stronger than the P− to P− interaction. (c) Both interactions become stronger when the salt concentration is 160 mM NaCl.

From Figure 3a, we see that for both P− on P− and P+ on P+ the MSDs are linear with the lag time ($\tau$), indicating a dominant fraction of the ensemble performs a purely diffusive motion. The log-log plot (inset) shows a slope of ~1.0 for both, reaffirming this observation.

However, the average diffusion coefficients extracted from the MSDs (Figure 3a) are quite different for the two cases of P− on P− and P+ on P+. A clearer picture emerges when we look closely into the MSD plots of each particle constituting the ensemble. We see that $P(n)$ for P− on P− is majorly peaked around n =1.0, indicating predominantly free diffusion, but $P(n)$ for P+ on P+ shows two peaks, one centered about $n = 1.0$ as before, and the other centered about $n = 0.15$ (Figure 3b), indicating that nearly 50% of the particles are undergoing subdiffusion. These results demonstrate the power of our technique, enabling us to identify subpopulations that otherwise would be masked by the ensemble average in bulk measurements. The bimodal distribution of the ensemble for P+ on P+ is reaffirmed in the $G(\Delta x)$ plots (Figure 3c). Here, $G(\Delta x)$ for P+ on P+ can be fitted with two Gaussians with variances 0.3 μm² and 10.5 μm², whereas $G(\Delta x)$ for P− on P− can be fitted approximately with a single Gaussian. Note that for P+ on P+, the nature of the $G(\Delta x)$ curve is distinctly non-Gaussian even though the ensemble time averaged MSD is linear in lag time. The result for P+ on P+ is a clear example of Fickian yet non-Gaussian diffusion.[41,42] This type of diffusion has been previously observed in particles diffusing on phospholipid tubules and in entangled actin filaments.[41] 'Diffusing diffusivity' that is the variable nature of the diffusion constant of each particle owing to the nature of intermittent forces acting on them allowing them to bind and unbind has been suggested as one possible reason behind this type of behavior.[43] In our system, due to the polyampholytic nature of the peptide sequences, weak, transient bonds are expected to form via ionic bridging between neighboring sequences.[12,14,44–47] Broken and re-formed ionic bridges will continuously give rise to 'hop and stick' trajectories and a considerable spread in the sticking time distributions. Similar reasons are likely behind the Fickian yet non-Gaussian diffusion in our system.

Direct measurements of the transport coefficients from the individual MSD plots (Figure 3d) show the two existing populations in the P+ on P+ system, with $P(\widetilde{D})$ peaked once at $\widetilde{D} =$ 0.1 $\mu m^2/s^n$ and also at $\widetilde{D}$ = 4.8 $\mu m^2/s^n$. In contrast, for P− on P− the first peak at $\widetilde{D}$ = 0.1 $\mu m^2/s^n$ is very small, almost at the level of the measurement error, with the main peak at about $\widetilde{D}$= 5.0 $\mu m^2/s^n$. The two-state nature of P+ on P+ therefore clearly arises from the single mutation of replacing the amino acid E with K. To rule-out inhomogeneous surface coverage as a compounding factor, we looked at the spatial distributions of both $\widetilde{D}$ and $n$ (Figure S5) and found that the variations in these two parameters are uniform throughout the imaged area (175 × 120 μm).

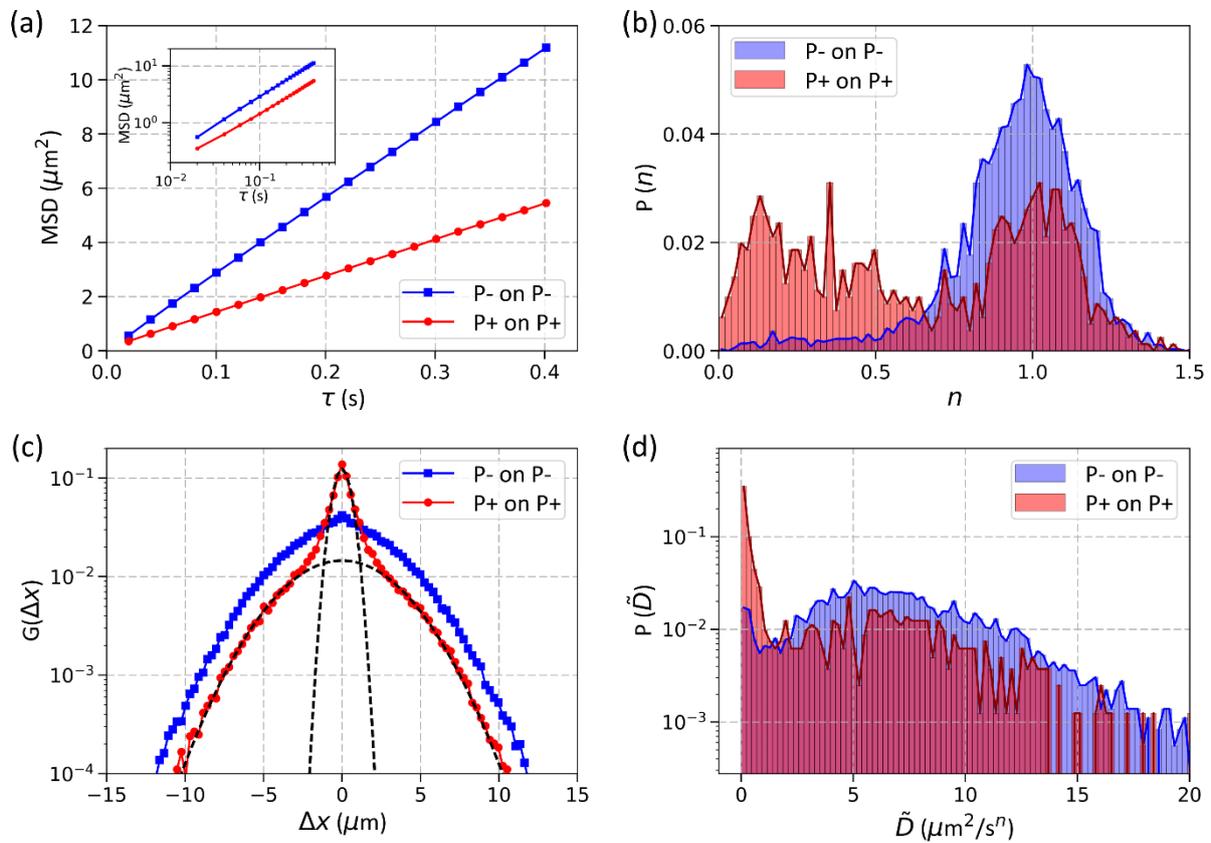

**Figure 3.** Diffusion of P− coated particles on P− coated glass and P+ coated particles on P+ coated glass. (a) Time ensemble averaged MSD plots. Inset shows the log-log plot of the same graph. The slope is nearly 1 for both the cases. (b) $P(n)$ plot shows a single peak for P− on P− and two distinct peaks for P+ on P+ indicating nearly 50% of the population undergoing subdiffusion in the latter. (c) $G(\Delta x)$ plot for P+ on P+ can be fitted to two

Gaussians, indicating the bimodal distribution of the ensemble. (d) $P(\widetilde{D})$ for P+ on P+ also shows two peaks. The number of particles analyzed for the given data were $N$ = 3617 for P− on P− and $N$ = 806 for P+ on P+.

An important handle to control weak peptide-peptide interactions, and in particular ionic bonds, is buffer salinity. An addition of monovalent salt screens the long-range repulsion between two peptides, allowing them to come close to each other. However, this, in turn, can promote the short-range attractions between the differently charged amino acid moieties on each peptide.[48–51] As a result, with the addition of salt, an increased number of bonds can form between the coated GNPs and coated surface, which will lead to slower diffusion. Indeed, this is evident from the decreasing slope of the time ensemble averaged MSD plots for P− on P− with salt in Figure 4a. We also note that the average diffusion exponent $n$ in this case decreases slightly from 0.99 to 0.92 as we increase the salt concentration from 0 mM to 160 mM NaCl. Additionally, the fraction of particles undergoing subdiffusion increases gradually with the addition of salt (Figure 4b). Fitting two Gaussians to the $G(\varDelta x)$ plot at 160 mM NaCl shows that there is a substantial deviation from Gaussian behavior (Figure 4c) for high $\varDelta x$ values. This deviation indicates that at these high salt concentrations, there is a wide distribution in the transport coefficients of the mobile GNPs leading to a departure from Gaussian tails in $G(\varDelta x)$. The transport coefficient distribution $P(\widetilde{D})$ shows that the fraction of slowly moving particles increases while the fraction of fast-moving particles decreases with an increase in salt concentration (Figure 4d). For P+ on P+, the effect of salt addition is even more drastic (Figure S6). Not only does $n$ change from 0.93 to 0.81 upon addition of 160 mM NaCl, but the ensemble statistics dramatically change. In particular, $P(n)$ shows a transition from ~50% of the ensemble being peaked near $n$ = 0.15 at low salt concentration to 81% of the ensemble peaked at $n$ = 0.15 at high salt concentrations. This is in contrast to the P− on P− case where 38% of the particles have a diffusion exponent about $n$ = 0.15 at 160 mM NaCl compared to 10% at 0 mM NaCl. The results for $G(\Delta x)$ and $P(\widetilde{D})$ indicate an even stronger effect of salt on the P+ on P+ interaction than on the P− on P− interaction (Figure S6).

Finally, we performed two additional experiments to evaluate the cross-interactions between P+ and P−. To this end, we study two systems, P+ on the GNPs and P− on the glass surface and vice versa. The main difference between the peptide grafting to GNPs and glass is that the connection to the GNPs is through a single immobile thiol-Au bond, but the connection to the glass surface is via a chain of a 10 kDa PEG molecule. Also, the grafting density on the GNP surface varies slightly between the two peptides, having the values of 0.48 and 0.68 nm$^{-2}$ for P− and P+ respectively.

If grafting geometry was unimportant, these two systems would have yielded similar results. Nonetheless, the different results obtained in these two systems suggest that molecular details can be substantial in altering the GNP diffusion (Figure 5). For the case of P− on P+, we observe an overall faster diffusion than for the reverse case. Even though the ensemble averaged MSDs are linear in time for both cases (Figure 5a), the fraction of particles undergoing subdiffusion is larger for P+ on P− (Figure 5b). From both $G(\Delta x)$ and $P(\widetilde{D})$, we find that the P− on P+ system has clearly a larger fraction of the particles with a higher transport coefficient in comparison to the P+ on P− system. Moreover, the effect of salt is different for the two systems. While the addition of 160 mM NaCl barely changes the diffusion parameters for P− on P+ (Figure S7), it has a drastic effect on the P+ on P− system, causing a significant increase in the fraction of slowly moving particles (Figure S8).

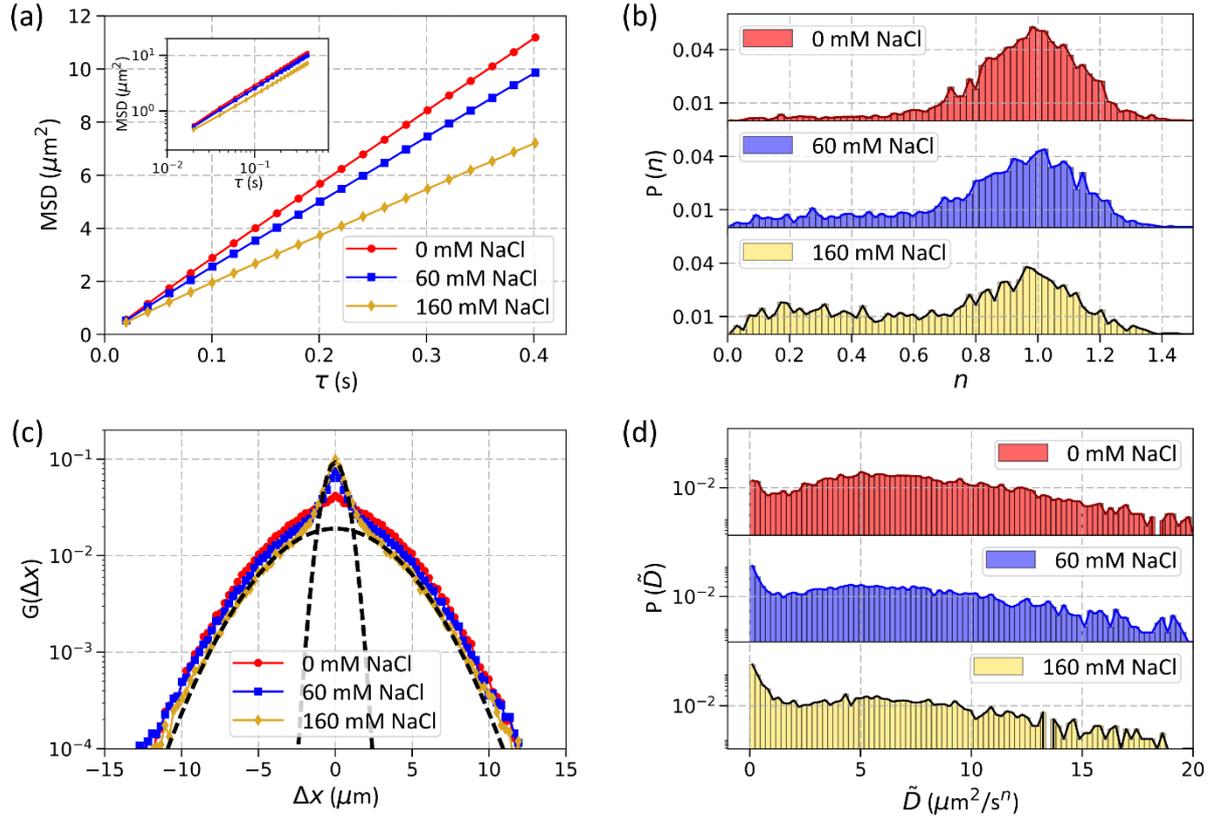

**Figure 4.** Diffusion of P− coated particles on P− coated glass at three different salt concentrations of 0 mM, 60 mM, and 160 mM NaCl. (a) Time-ensemble averaged MSD plots. Inset shows the log-log plot of the same graph. Note the decrease in slope in both the linear and log-log plot with increasing salt concentrations indicating both slower diffusion and a shift towards subdiffusion. (b) $P(n)$ calculated from the log-log MSD plots of each member of the ensemble. (c) $G(\Delta x)$ distribution (d) $P(\widetilde{D})$ distribution. The number of particles analyzed for the given data were $N = 3617$, $N = 2820$ and $N = 2060$ for 0 mM NaCl, 60 mM NaCl and 160 mM NaCl respectively.

We now have the following overall trend regarding the mobility of the various coated GNPs near the coated surfaces. GNPs in the P+ on P+ system seem to be the least mobile with a large fraction of particles undergoing subdiffusion. GNPs in the P− on P− system are the most mobile, exhibiting nearly free diffusion. In between these extremes, we have the P+ on the GNPs and P− on the surface, which is less mobile than the opposite arrangement. This is observed clearly from the progressively increasing fraction of subdiffusing particles ($\varphi_s$) with $n \leq 0.5$ as one goes from P− on P−, to P− on P+, P+ on P− and P+ on P+ respectively as shown in Figure 6a. Another measure of the binding strength of GNPs to the glass surface is the equilibrium dissociation constant $K_d$ for binding of GNPs to the glass surface. To illustrate the differences in interaction strengths between the four peptide systems, we calculate $K_d$ in an analogous fashion to the approach used by Yoo *et al.* for weakly interacting fluorescence tagged proteins.[52] In our model, we assume the GNPs to be effectively 'molecules' binding and unbinding to the glass surface. We assume that the glass surface is fully covered with binding sites. Here, $K_d$ is given as $K_d \sim \frac{K_{OFF}}{K_{ON}} \sim A exp(\frac{\Delta G_B}{k_B T})$, where $K_{ON}$ is the rate constant of binding events given by $K_{ON} = \frac{f}{[GNP]}$, $f$ being the frequency of the binding events and [GNP] the measured concentration of GNPs diffusing on the surface, $K_{OFF} = \frac{1}{\langle t_{st} \rangle}$ is the rate constant of unbinding events and $\Delta G_B$ is the binding free energy per GNP. Since all the experiments were conducted at the same temperature, $K_d$ should reflect the change in binding free energy between the four peptide systems. A plot of $K_d$ for the four peptide systems showed an overall trend of decrease from P− on P− to P+ on P+ indicating a higher interaction strength for the later (Figure 6b), since the binding free energy $\Delta G_B$ is negative. The values of $K_d$ calculated for the four systems were much lower than that in ref. 52. This implies an effective stronger interaction strength resulting from the formation of multiple weak bonds between a GNP and the glass surface as compared to the weak bonds formed between two protein molecules in ref. 52. For DNA coated beads diffusing on a DNA coated surface with intermittent binding, Xu *et al.*[36] estimate the binding free energy of each bead as $\Delta G_B = -RT ln\left(\left(1 + je^{-\frac{\Delta G_{tether}}{RT}}\right)^{N_b} - 1\right)$ where $N_b$ is the maximum number of bonds that can be formed in the bond area, $j$ is the the number of bonds that opposing DNA sticky ends can reach and $\Delta G_{tether}$ is the binding free energy of each single bond. It is obvious from this that a small increase in $N_b$ would result a large decrease in $K_d$. It should be noted here that our estimation of $K_d$ is extremely simplistic. Each bond between a GNP and the peptide coated glass surface consists of several peptide-peptide bond pairs. To accurately estimate the value of $K_d$ the average number of bonds forming between the GNP and the glass surface in each sticking event should be measured. We estimate an order of at least a hundred bonds forming between the GNP and glass, taking into account the geometry of the contact and the grafting density, but refer for future work to measure the exact value. We expect, therefore, several orders of magnitude increase in the measured binding energy as compared to the measured value for a single protein.[52]

There are several differences between the two systems of P− on P+ and P+ on P− leading to the observed asymmetric results. First, as noted before, the grafting density of P− is somewhat lower than P+. Due to the highly curved surface of the spherical GNPs, such changes in grafting can result with lower fraction of P− peptide molecules interacting with the surface. Second, the overall charge of P− is larger than that of P+ and thus can lead to more stretched conformations of P−. This will have a more pronounced effect when P− is grafted to the GNP surface. Stretched conformations of the peptide on the GNP surface will again lead to a smaller number of peptides interacting with the peptides on the glass. Finally, the presence of the 10 kDa PEG linker molecule between the glass surface and the attached peptide provides more freedom to the peptides on the glass to rearrange in space and interact with its neighboring peptides compared to the one attached to the GNPs. All of the above details support the picture that when P+ is attached to the glass surface, it bonds more with adjacent molecules on the surface and interacts less with the P− peptides on the GNPs. In the opposite case, when the P+ is attached to the GNPs, it has less opportunity to self-interact and is more available to bond with the P− peptides on the glass surface, leading to a much stronger probability of attachment. A schematic showing the different degrees of interaction for the four GNP peptide systems is given in Figure 6c-f. Note the formation of bonds between the neighbouring P+ peptides (shown in red) on the glass while the P− peptides (shown in blue) on the glass have more stretched conformations. These simple qualitative arguments underline the importance of geometry in the observed peptide-peptide interactions. Furthermore, to have a preliminary evaluation regarding the dynamics of our system, we calculated the scaling of the probability distribution of sticking time, $P(t_{st})$ in all the four cases. For continuous time random walk (CTRW) problems, a scaling law of the form $P(t_{st}) \sim \frac{1}{t_{st}^{1+\alpha}}$ is expected where $0 < \alpha < 1$ and $MSD \sim t_{st}^{\alpha}$. Xu *et al.* used the scaling behavior of $P(t_{st})$ to study the diffusive motion of DNA coated particles on a DNA coated surface.[36] By fitting a straight line to the log-log plot of $P(t_{st})$ vs $t_{st}$ for all the four peptide systems, we found that the value of $\alpha$ for the four peptide systems were: $\alpha$ = 2.4 ± 0.2 for P− on P−, $\alpha$ = 1.7 ± 0.1 for P− on P+, $\alpha$ = 1.0 ± 0.1 for P+ on P− and $\alpha$ = 1.2 ± 0.1 for P+ on P+. Therefore, instead of showing CTRW behavior, the tracer particles diffused almost normally with the exception of the most charged system P− on P−. This could be a result of the weaker interactions between the peptides as compared to complementary DNA strands, or of the lack of sufficiently long trajectories due to the small size of the GNPs. Further theoretical and experimental analysis of this effect will be detailed in future communications supported by measurements involving temperature and force control.

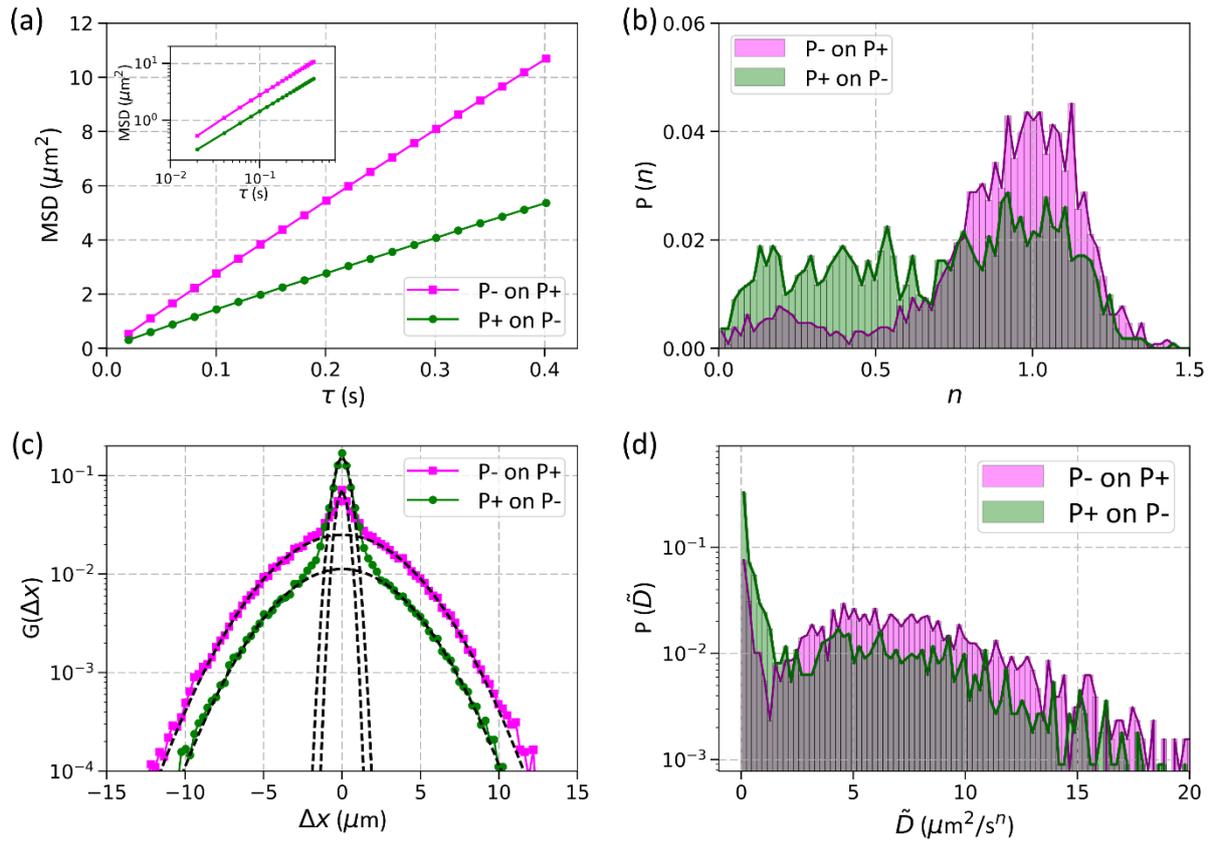

**Figure 5.** Diffusion of P− coated particles on P+ coated glass and P+ coated particles on P− coated glass. (a) Time- ensemble averaged MSD plots. Inset shows the log-log plot of the same graph. The slope is nearly 1 for both the cases. (b) $P(n)$ calculated from the log-log MSD plots of each individual member of the ensemble. (c) $G(\Delta x)$ distribution (d) $P(\widetilde{D})$ distribution. The number of particles analyzed for the given data were $N = 1284$ for P− on P+ and $N = 1113$ for P+ on P−.

To conclude, weak and transient peptide-peptide interactions were studied by exploring the diffusion of peptide coated GNPs on peptide coated glass surfaces. This effective method was capable of detecting the effect of a single mutation in the amino acid sequence. We found that replacement of one single glutamic acid residue with a lysine residue leads to very different diffusive behaviors characterized by widely different MSDs, distributions of diffusion exponents, probe particle displacements and transport coefficient distributions. Salt was found to be an effective handle to control the weak interactions; however, the extent of that control also depended upon the nature of the peptide: the peptide with the overall lesser charge responded more strongly to changed salt concentrations. The detailed molecular arrangements also proved to be an important influencing factor as evidenced by the different diffusive behaviors obtained by coating the GNPs and the glass surfaces with two different peptides and then reversing them. Our technique proved to be an efficient method for comparative evaluation of the nature of different peptides and their interactions.

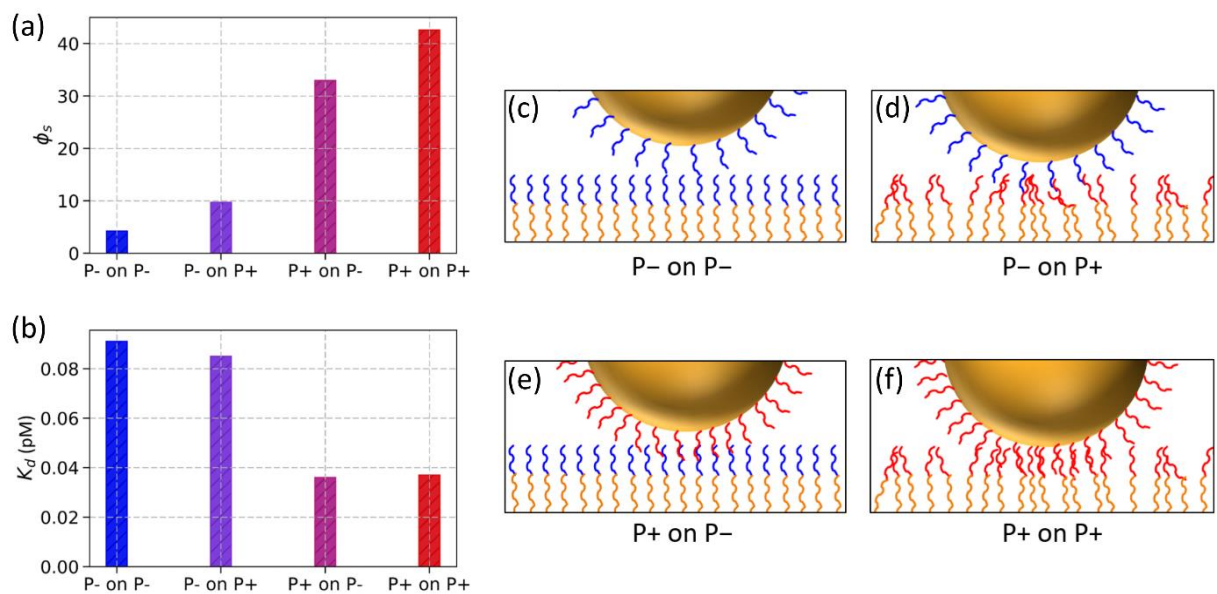

**Figure 6.** Measures for the interaction strengths in the four systems: P− on P−, P− on P+, P+ on P− and P+ on P+. (a) fraction of subdiffusing particles ($\phi_s$) with $n \leq 0.5$. Note the progressive increase as we go from P− on P− to P+ on P+ via the two cross-interaction cases. (b) Equilibrium dissociation constant for the four different systems. $K_d$ has an overall decrease as we go from P− on P− to P+ on P+, indicating an overall increase in the interaction strength. A schematic of our proposed model for the four peptide systems is shown from (c)-(f) with P− represented by blue colour and P+ represented by red colour. Note the increased number of bond formation and consequently higher interaction strengths for GNPs coated with the P+ peptide.

Our technique can easily be modified to work at high throughput conditions by integrating it into a microfluidic device. Moreover, the range of interaction strengths that can be quantified by this method can be extended significantly by changing the number of available bonds by changing the size of the GNPs. Even though we demonstrated our technique using synthetic peptides, it can be applied to study and quantify relative strengths of weak interactions in many biological systems.

Methods:

Circular dichroism (CD) measurements

Far-UV-CD spectra of 10 μM peptides were measured in 5 mM TRIS pH 8.0 with a 5 mm cuvette at 25°C using a Chirascan Circular Dichroism Spectrophotometer (Applied Photophysics, UK). Each scan was recorded over the range of 190–260 nm, bandwidth 1 nm, with average time per point of 0.5 sec. For each sample, five scans were measured and averaged.

Preparation of GNPs coated with peptides

Synthesis and purification of peptides were performed by the Blavatnik Center for Drug Discovery at the Tel Aviv University. Peptides were identified using mass spectrometry and were purified over 95% by HPLC. Peptide sequences were design to have Cysteine residue at the N-termini to allow formation of thiol-gold bond with the GNPs. 40 nm GNPs stabilized by citrate, were purchased from BBI Solutions. At first ~100 μM peptide was added to ~0.15 nM GNPs for 1 hour at room temperature. Next, TRIS buffer of pH 8.0 was added to a final concentration of 20 mM. The solution was left at mild tilt overnight. The next day excess peptides were removed by repeated (four times) filtration using Amicon 100 kDa ultra-0.5 centrifugal filter units (Millipore, UFC510024) for 2 min at 5000g. After each filtration step, the solution was diluted five times with 20 mM TRIS buffer of pH 8.0. The final solution was stored at 4°C until further measurements.

Measuring peptide grafting density on the GNPs

Concentration of the GNPs was determined by analyzing the UV-VIS absorbance at 500 nm. Fluorescence spectra of the peptide coated GNPs containing Tryptophan were measured using a FL3-11 Spectrofluorometer (HORIBA, UK). Measurements were done in a 5 mm quartz cuvette at GNP concentrations of ~ 6 pM in 10 mM TRIS buffer pH 8.0 at room temperature. The excitation wavelength was 280 nm (bandwidth 2 nm) and the emission spectrum was recorded over the range of 320–500 nm (bandwidth 5.0 nm). Peptide concentration was determined by using N-Acetyl-L-tryptophanamide (NATA) as a reference. NATA concentration was measured by absorbance at 280 nm and the intensities of the fluorescence emission of both peptide and NATA were normalized to determine the peptide concentration. Buffer and background signals were routinely measured and subtracted. Grafting density was calculated by dividing the measured concentration of the peptides with the area of the total measured concentration of the GNPs, assuming a uniform distribution of peptides on the spherical GNPs.

In order to ensure a good peptide coverage on the GNP surfaces, we ensured that the surface was saturated with the peptides. GNPs were functionalized with different initial peptide concentrations in the range of 1uM – 200uM (see Figure S1). It was found that the grafting saturated at ~0.48 peptide/nm$^2$ for the P− peptide. Henceforth GNPs coated with the same grafting density were used for all the diffusion experiments. To exclude error in grafting density measurements due to fluorescence quenching of the Tryptophan residue under saturation coverage, a control experiment was done. 10mM β-mercaptoethanol was added to the measured solution to break the Au-thiol bonds and bring the peptides directly into the solution. However, the increase in fluorescence was less than 5% indicating no substantial quenching of fluorescence at saturation grafting density.

Preparation of peptide coated glass cover-slips

Glass cover slips of (22 × 22 mm) and (24 × 50 mm) obtained from Deckglässer were at first cleaned thoroughly by washing in 2% Hellmanex and deionized water, and subsequent ultrasonication for 30 min in 2:1 methanol and water solution. Then they were dried in the oven at 80 °C for 1 hour and oxygen plasma cleaned for 5 minutes just before functionalization with PEG/peptides. The plasma cleaned 22 × 22 mm cover slips were coated with silane-PEG-maleimide of molecular weight 10 kDa obtained from Nanocs. 10 mg of silane-PEG-maleimide was added to 450 μL dehydrated ethanol (purity > 98%) and 19 μL deionized water. 100 μl of this solution was pipetted onto one glass cover slip and the plasma cleaned surface of another cover slip was slowly and carefully slipped onto this drop in order to form a liquid film sandwiched between the two cover slips. The coating was done for two hours and then the cover slips were rinsed with deionized water and dried under $N_2$ gas. The 24 × 50 mm cover slips were coated with PEG-silane (molecular weight 5 kDa, obtained from Nanocs) via a similar procedure. In this case, a solution containing 10 mg of PEG-silane in 650 μL ethanol, 40 μL water and 6.5 μL acetic acid was made and 150 μL of this solution was again pipetted onto one glass cover slip and sandwiched with another. The PEG-silane coating of the 24 × 50 mm cover slips prevents the peptide coated GNPs from sticking to the surface. On the other hand, the 22 ×22 mm cover slips are coated with silane-PEG-maleimide where the silane gets covalently attached to the glass surface, and the maleimide forms a thiol-maleimide bond with cysteine containing peptides under ambient conditions. For peptide coating of these glass cover slips, 150 μL of a 0.2 mg/mL peptide solution in 20 mM TRIS buffer of pH 7.5 was taken and pipetted onto a previously silane-PEG-maleimide coated cover slip. This was sandwiched gently with another similar cover slip taking care that no liquid leaks out from the space between the two cover-slips. The peptide functionalization was allowed to take place overnight. Subsequently the peptide coated cover-slips were washed with deionized water and dried under $N_2$.

Preparation of the sample chamber

To prepare the sample chamber for the diffusion experiments, 10 μL of ~0.6 pM peptide coated GNPs in 20 mM TRIS buffer of pH 7.5 was injected onto to a peptide coated cover slip and sandwiched with a 24 × 50 mm PEG-silane coated cover slip (see Figure 1a) on top. After a few seconds the whole sample chamber was inverted and glued together with UV-curable glue.

Imaging and particle tracking

Imaging was done in dark-field mode with an Olympus IX 71 microscope with 20x/40x objectives. Images were recorded at 50 fps and subsequent post-processing was done in ImageJ before analyzing particle trajectories using Trackpy,[53] a Python based implementation of the Crocker-Grier algorithm.[54] We enabled several filters to select 'good' particle trajectories including a) selection of particles within a given size window to eliminate clusters or other spurious bright spots b) selecting sufficiently long trajectories containing at least thrice the number of points taken to calculate the MSD plots c) excluding particles which are completely stuck from the beginning of the experiment.

ASSOCIATED CONTENT

**Supporting Information**. The following files are available free of charge. Figures showing fluorescence measurement of grafting density on peptide coated GNPs, AFM phase and topography images of peptide and PEG coated glass surfaces, transport coefficient distributions of GNPs diffusing on a series of glass surfaces with different peptide grafting densities, mobile, immobile, fast and slow fractions among diffusing particles for the two cases of P− on P− and P+ on P+, 2D colour maps of transport coefficients and diffusion exponents for P− on P− and P+ on P+, and the effect of salt on different characteristic parameters of diffusion for P+ on P+, P− on P+ and P+ on P−. (PDF)

AUTHOR INFORMATION

**Corresponding Author**

*E-mail: roy@tauex.tau.ac.il

*E-mail: roichman@tauex.tau.ac.il

**Author Contributions**

IC prepared the glass surfaces, assembled the sample chambers, and carried out the particle tracking experiments and analysis. GR and RA designed the sequences. GR functionalized the GNPs with peptides and measured grafting densities. RA conceived the basic concept of the project. RB and YR lead the project. All the authors designed the experiments and participated in writing the paper.

**Funding Sources**

This work was supported by the Israel Science foundation (ISF) grants number 453/17, 550/15 and 988/17, and the United States - Israel Binational Science Foundation (201696).

**Notes**

The authors declare no competing financial interest.

ACKNOWLEDGMENT

IC acknowledges the PBC Post-doctoral fellowship of the Council for Higher Education of Israel. We acknowledge the helpful discussions with Fernando Patolsky, Yacov Kantor, Micha Kornreich and Guy Jacoby.

TABLE OF CONTENTS FIGURE

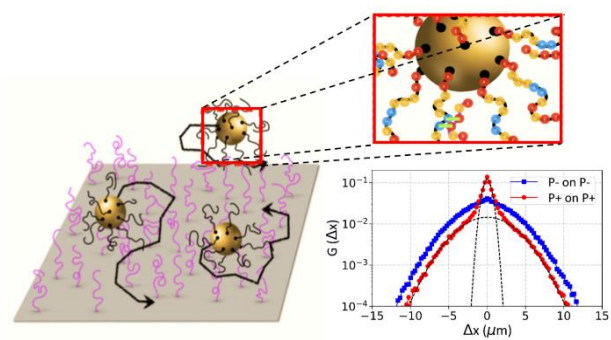

Supporting information

# Nanoparticle mobility over a surface as a probe for weak transient disordered peptide-peptide interactions

*Indrani Chakraborty [†], Gil Rahamim [‡], Ram Avinery [‡], Yael Roichman\* [†,‡] and Roy Beck\* [‡]*

[†]School of Chemistry, Tel Aviv University, Tel Aviv 6997801, Israel

[‡]School of Physics and Astronomy, Tel Aviv University, Tel Aviv 6997801, Israel

\*E-mail: roy@tauex.tau.ac.il

\*E-mail: roichman@tauex.tau.ac.il

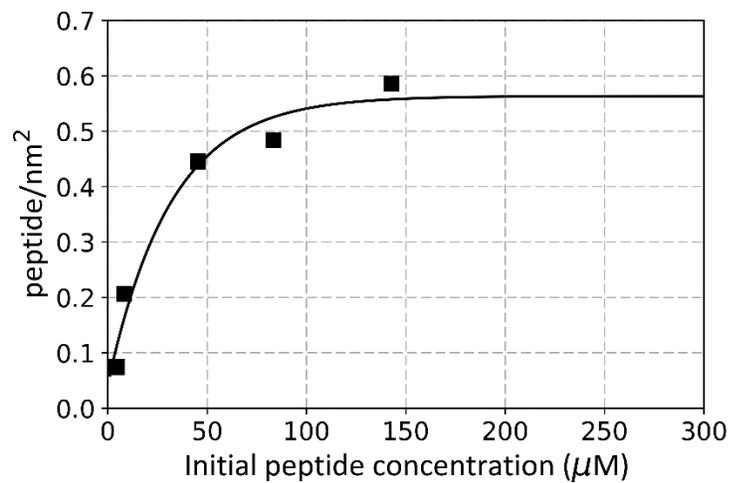

Figure S1. Measurement of grafting density of the P− peptide on the GNPs as a function of the initial peptide concentration using fluorescence of the tryptophan residue. P+ showed a similar behavior.

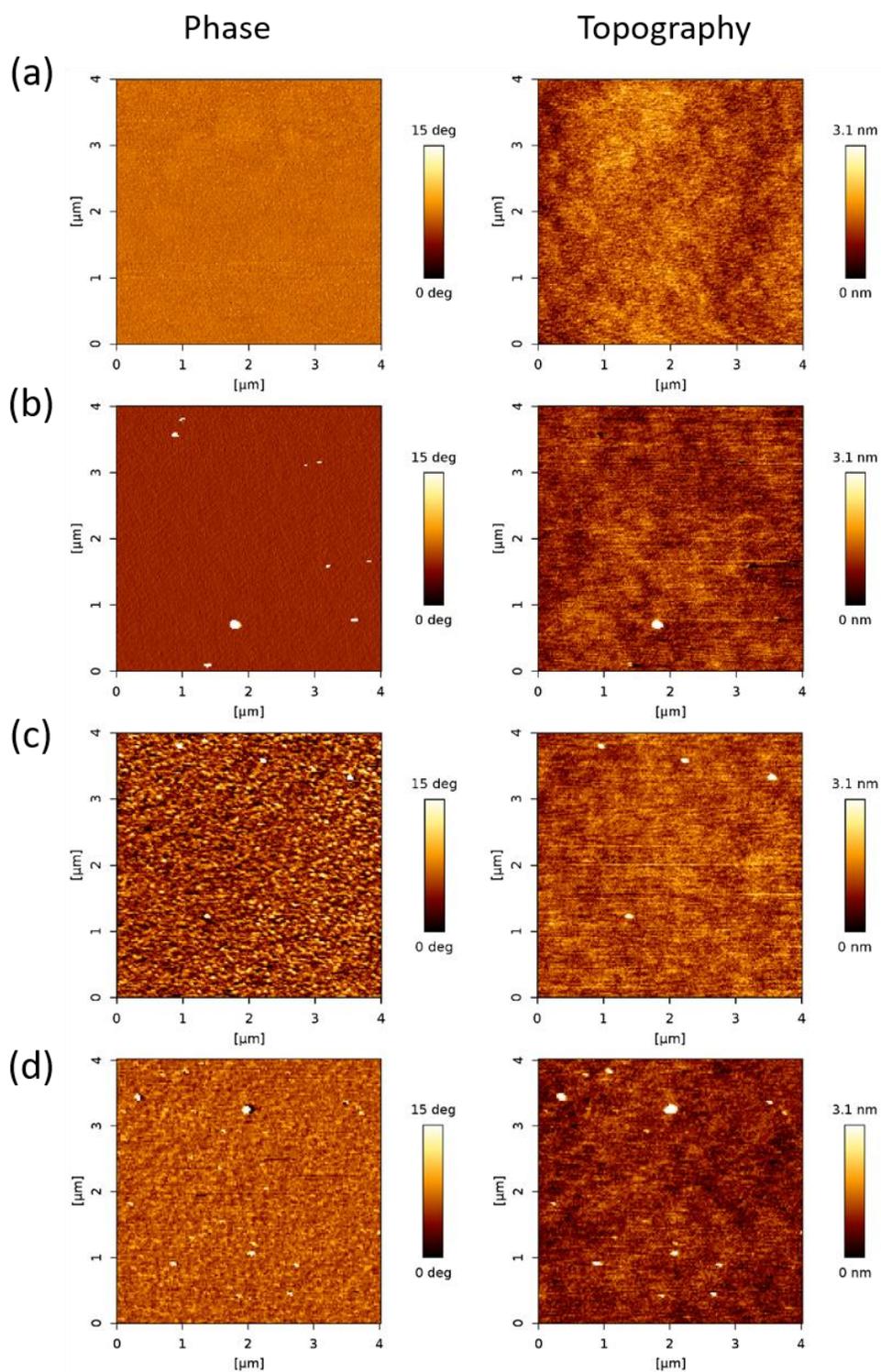

Figure S2. AFM phase and topography images of a) clean glass surface b) glass surface coated only with silane-PEG-maleimide c) glass surface coated with 50 μM initial concentration of peptide P− and d) glass surface coated with 200 μM initial concentration of peptide P−.

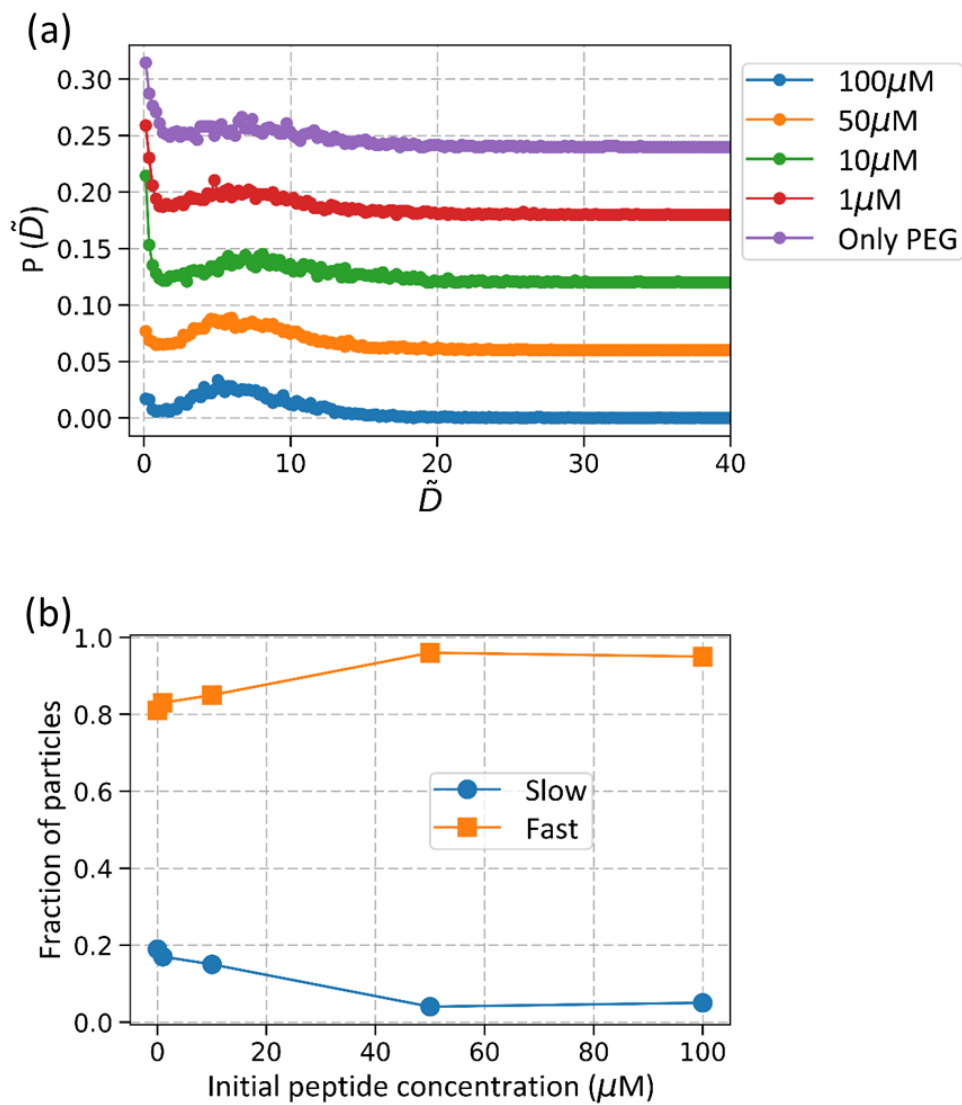

Figure S3: (a) Probability distribution $P(\widetilde{D})$ of the transport coefficient $(\widetilde{D})$ of P− peptide functionalized GNPs diffusing on glasses coated with different initial concentrations of the peptide P−. Note the decrease in the fraction of slowly moving particles (peaked around $\widetilde{D} = 0$ $\mu m^2/s^n$) and increase in the fraction of fast moving particles (peaked around $\widetilde{D} = 6$ $\mu m^2/s^n$) with increasing peptide concentration and subsequent saturation around 100 µM. (b) shows the fraction of slow particles ($\widetilde{D} < 1$ $\mu m^2/s^n$) and fast particles ($\widetilde{D} > 1$ $\mu m^2/s^n$) as a function of the initial peptide concentration.

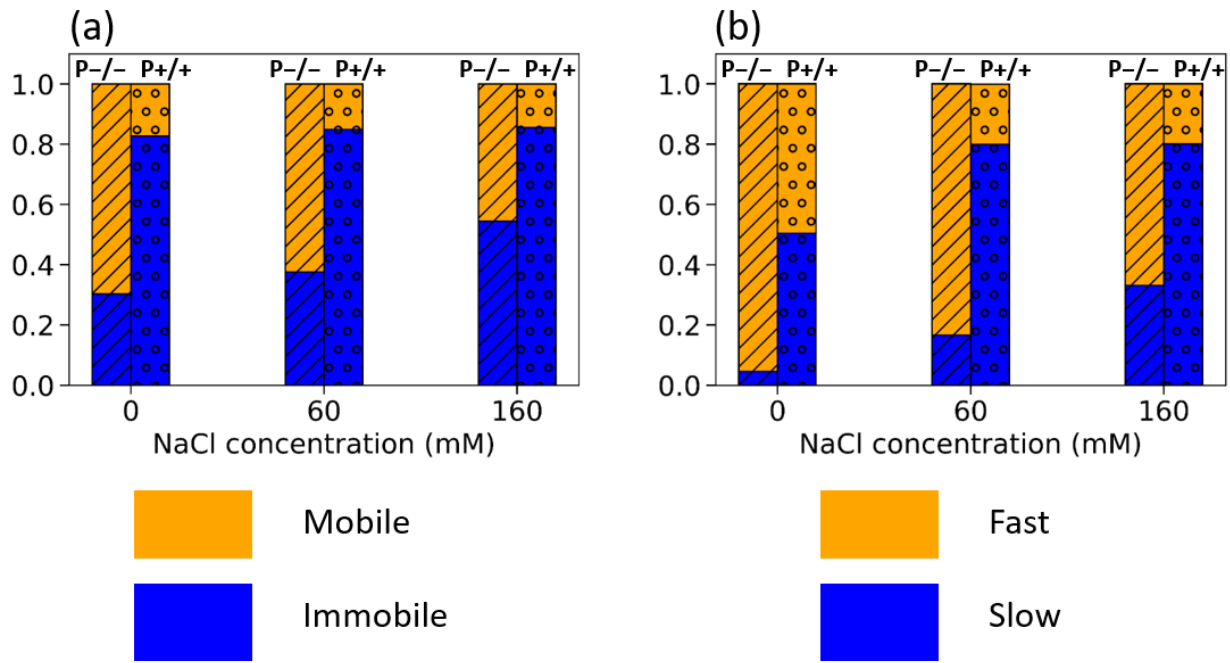

Figure S4. The difference in the diffusive behaviours of P− on P− and P+ on P+. Both (a) the fraction of mobile and immobile particles and (b) the fraction of slow and fast particles ('slow' for particles with $\widetilde{D} < 1\ \mu m^2/s^n$, 'fast' for particles with $\widetilde{D} > 1\ \mu m^2/s^n$ indicate that P+ on P+ has stronger interactions than P− on P−.

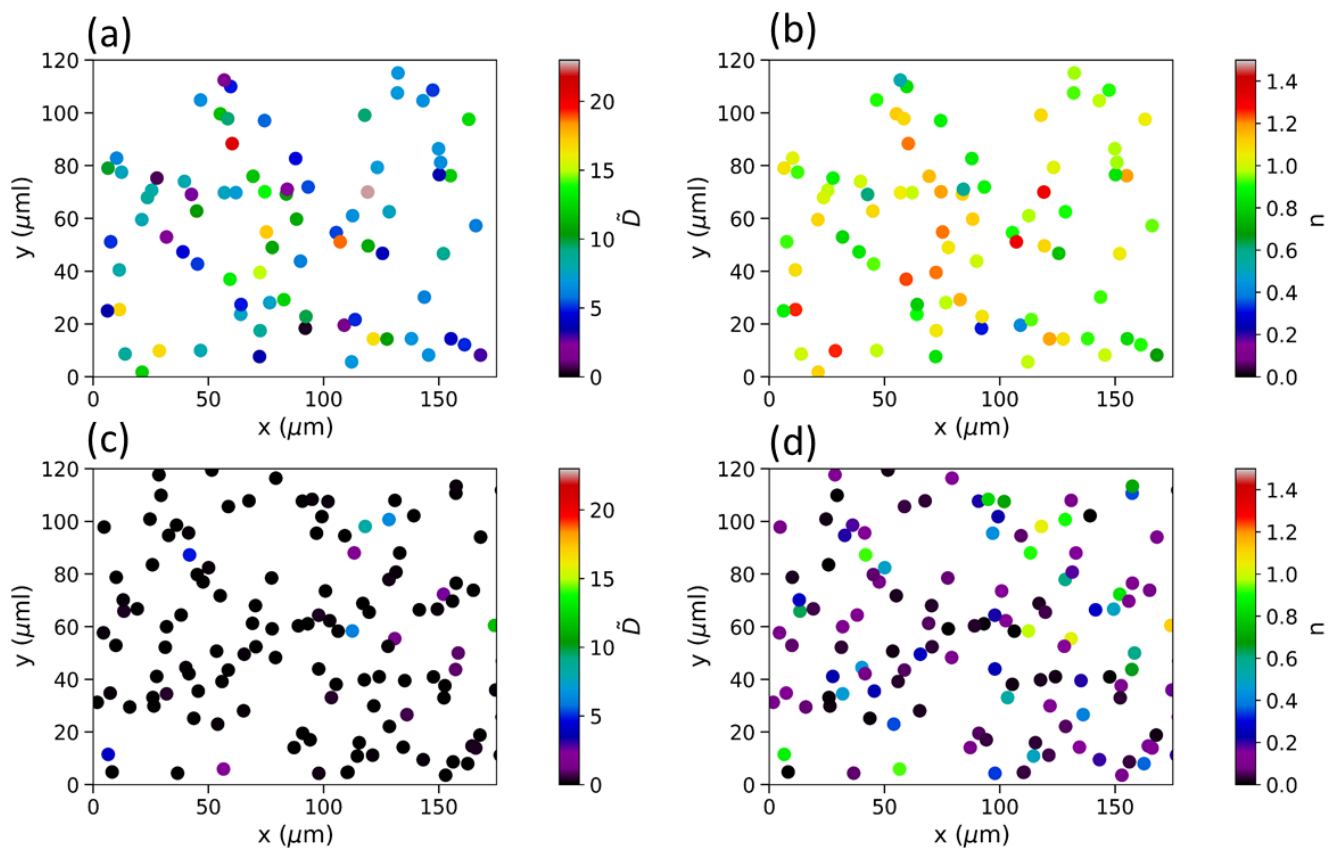

Figure S5. 2D colourmap of peptide coated GNPs diffusing on peptide coated glass surfaces as a function of their $\widetilde{D}$ and $n$ values. Warmer colors indicate higher transport coefficients or $n$ values. (a) and (b) are the $\widetilde{D}$ and $n$ value maps of P− on P− and (c) and (d) are the same for P+ on P+ respectively over an area of 175 × 120 μm.

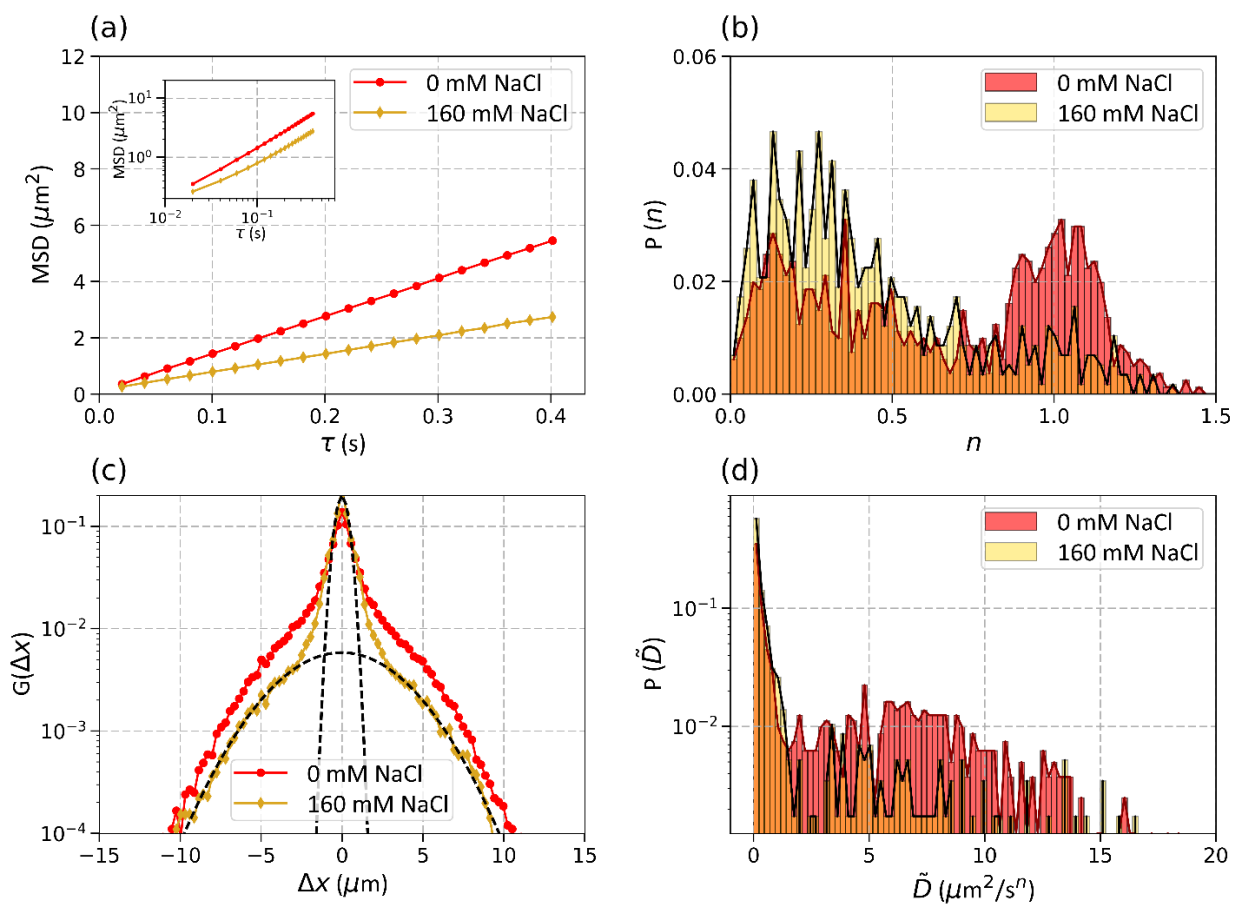

Figure S6. Diffusion of P+ coated GNPs on P+ coated glass at two different salt concentrations of 0 mM and 160 mM NaCl. (a) Ensemble averaged MSD plots. Inset shows the log-log plot of the same graph. (b) P($n$) calculated from the log-log MSD plots of each individual member of the ensemble. (c) G($\Delta x$) distribution (d) P($\widetilde{D}$) distribution. The number of particles analyzed for the given data were N = 806 for 0 mM NaCl and $N$ = 579 for 160 mM NaCl.

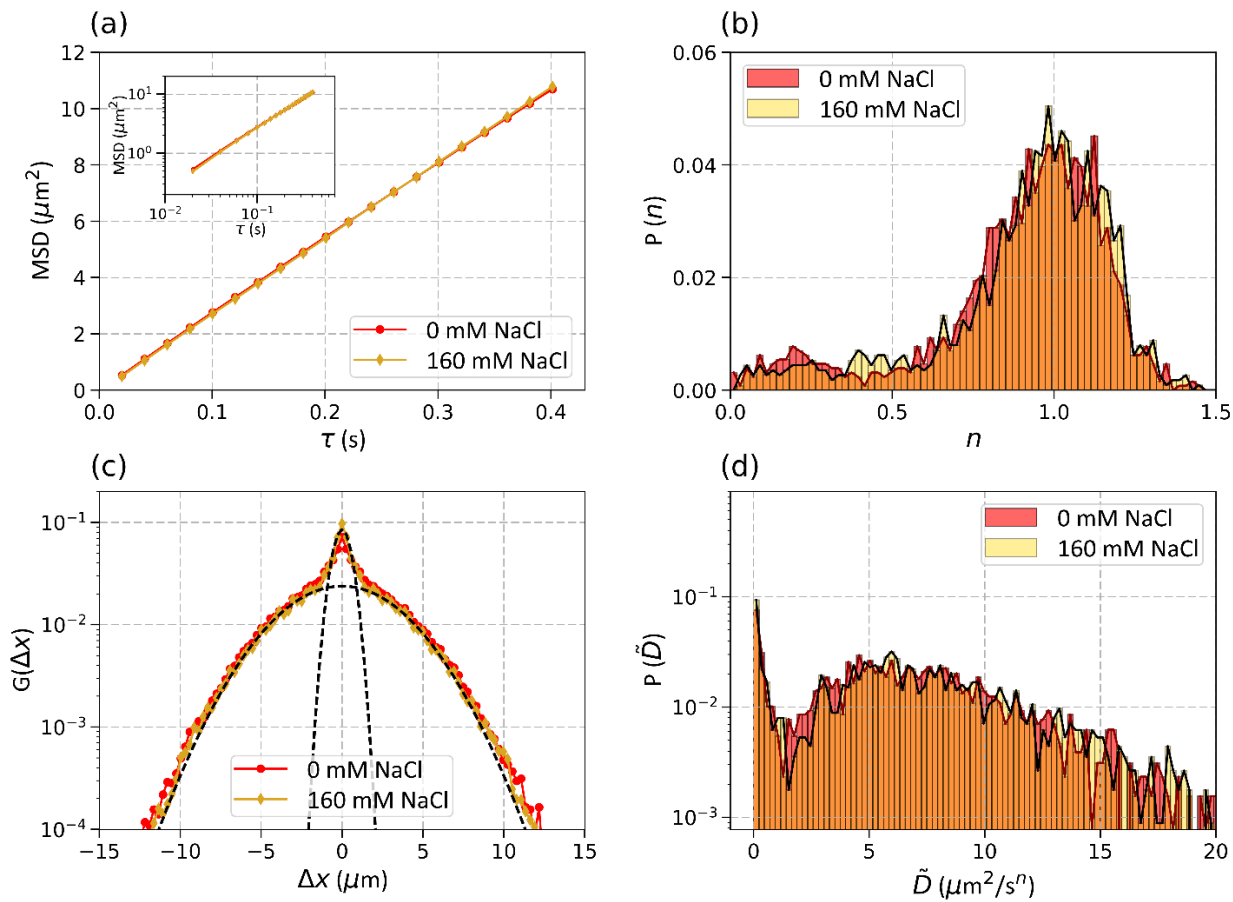

Figure S7. Diffusion of P− coated GNPs on P+ coated glass at two different salt concentrations of 0 mM and 160 mM NaCl. (a) Ensemble averaged MSD plots. Inset shows the log-log plot of the same graph. (b) P($n$) calculated from the log-log MSD plots of each individual member of the ensemble. (c) G($\Delta x$) distribution (d) P($\tilde{D}$) distribution. The number of particles analyzed for the given data were $N$ = 1284 for 0 mM NaCl and $N$ = 1130 for 160 mM NaCl.

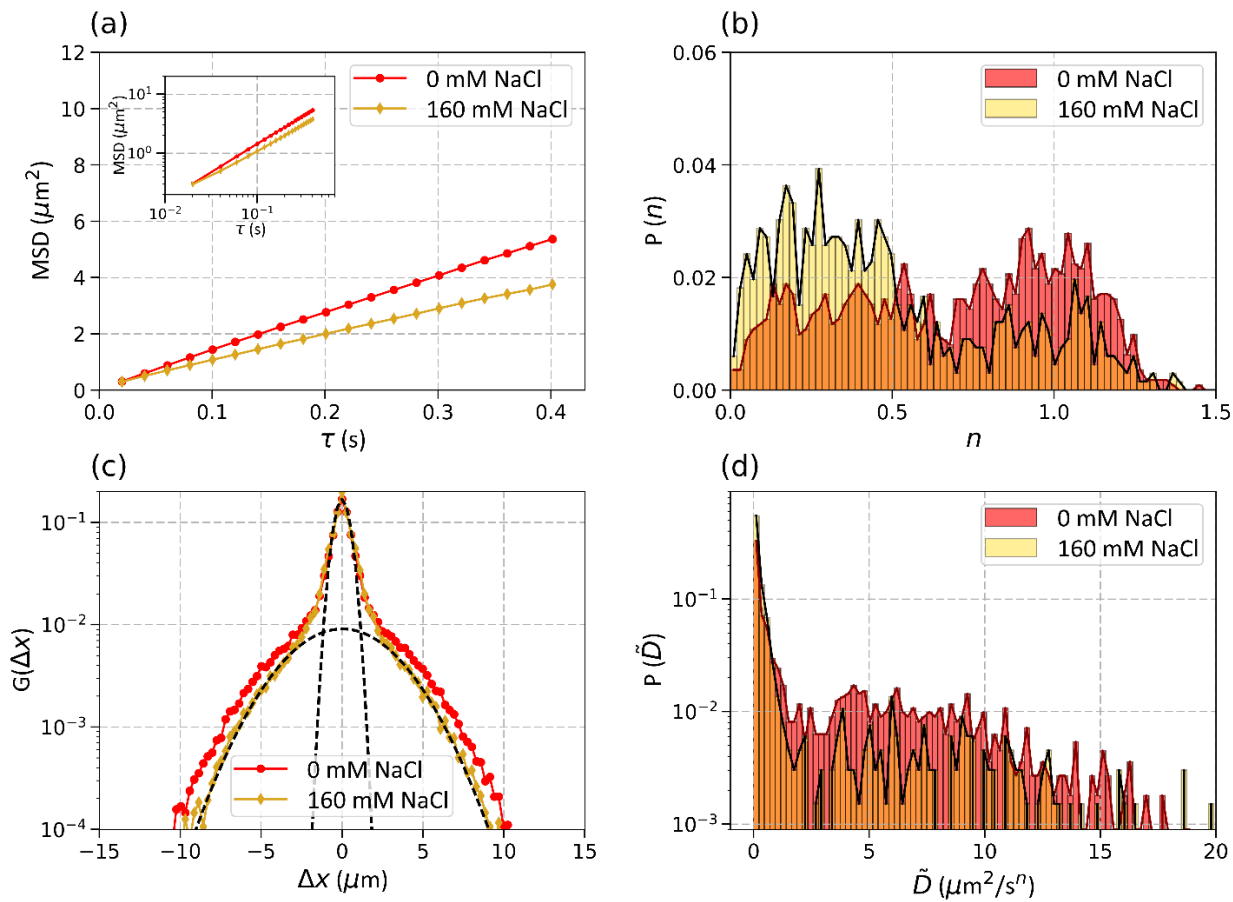

Figure S8. Diffusion of P+ coated GNPs on P− coated glass at two different salt concentrations of 0 mM and 160 mM NaCl. (a) Ensemble averaged MSD plots. Inset shows the log-log plot of the same graph. (b) P($n$) calculated from the log-log MSD plots of each individual member of the ensemble. (c) G($\Delta x$) distribution (d) P($\widetilde{D}$) distribution. The number of particles analyzed for the given data were $N$ = 1113 for 0 mM NaCl and $N$ = 661 for 160 mM NaCl.